\documentclass[preprint,showpacs,preprintnumbers,amsmath,amssymb,aps,floatfix]{revtex4}
\usepackage{graphicx}
\usepackage{dcolumn}
\usepackage{epsf}
\def\LL{\left\langle}	
\def\RR{\right\rangle}	
\def\LP{\left(}		
\def\RP{\right)}	
\def\LB{\left\{}	
\def\RB{\right\}}	
\def\PAR#1#2{ \frac{\partial #1}{\partial #2} }

\newcommand{\BE}{\begin{displaymath}}
\newcommand{\EE}{\end{displaymath}}
\newcommand{\BNE}{\begin{equation}}
\newcommand{\ENE}{\end{equation}}
\newcommand{\BEA}{\begin{eqnarray}}
\newcommand{\EEA}{\nonumber\end{eqnarray}}
\newcommand{\EL}{\nonumber\\}
\newcommand{\la}[1]{\label{#1}}

\newcommand{\OL}[1]{\overline{#1}\ } 
\newcommand{\OLL}[1]{\overline{\overline{#1}}\ } 
\newcommand{\OON}{\frac{1}{N}} 
\newcommand{\OOX}[1]{\frac{1}{#1}} 
\newcommand{\deltabar}{\OL{\delta}}
\newcommand{\ip}[0]{{i^\prime}} 
\newcommand{\jp}[0]{{j^\prime}} 
\newcommand{\kp}[0]{{k^\prime}} 
\newcommand{\lp}[0]{{l^\prime}} 
\renewcommand{\mp}[0]{{m^\prime}} 
\newcommand{\np}[0]{{n^\prime}} 
\newcommand{\pp}[0]{{p^\prime}} 
\newcommand{\rp}[0]{{r^\prime}} 
\newcommand{\is}[0]{{i^*}} 
\newcommand{\js}[0]{{j^*}} 
\newcommand{\dof}{{\it d}} 

\begin{document}

\title{Sample size effects in multivariate fitting of correlated data}

\author{D. Toussaint and W. Freeman}
\affiliation{Department of Physics, University of Arizona, Tucson, AZ 85721, USA}

\date{\today}

\begin{abstract}
A common problem in analysis of experiments or in lattice QCD
simulations is fitting a parameterized model to the average over
a number of samples of correlated data values.   If the number of
samples is not infinite, estimates of the variance of the parameters
(``error bars'') and of the goodness of fit are affected.  We illustrate
these problems with numerical simulations, and calculate
approximate corrections to the variance of the parameters
for estimates made in the standard way from derivatives
of the parameters' probability distribution as well as from jackknife
and bootstrap estimates.
\end{abstract}

\maketitle














%

\section{Introduction}

\begin{figure}[tbh]
\epsfxsize=5.5in
\epsfbox[0 0 4096 4096]{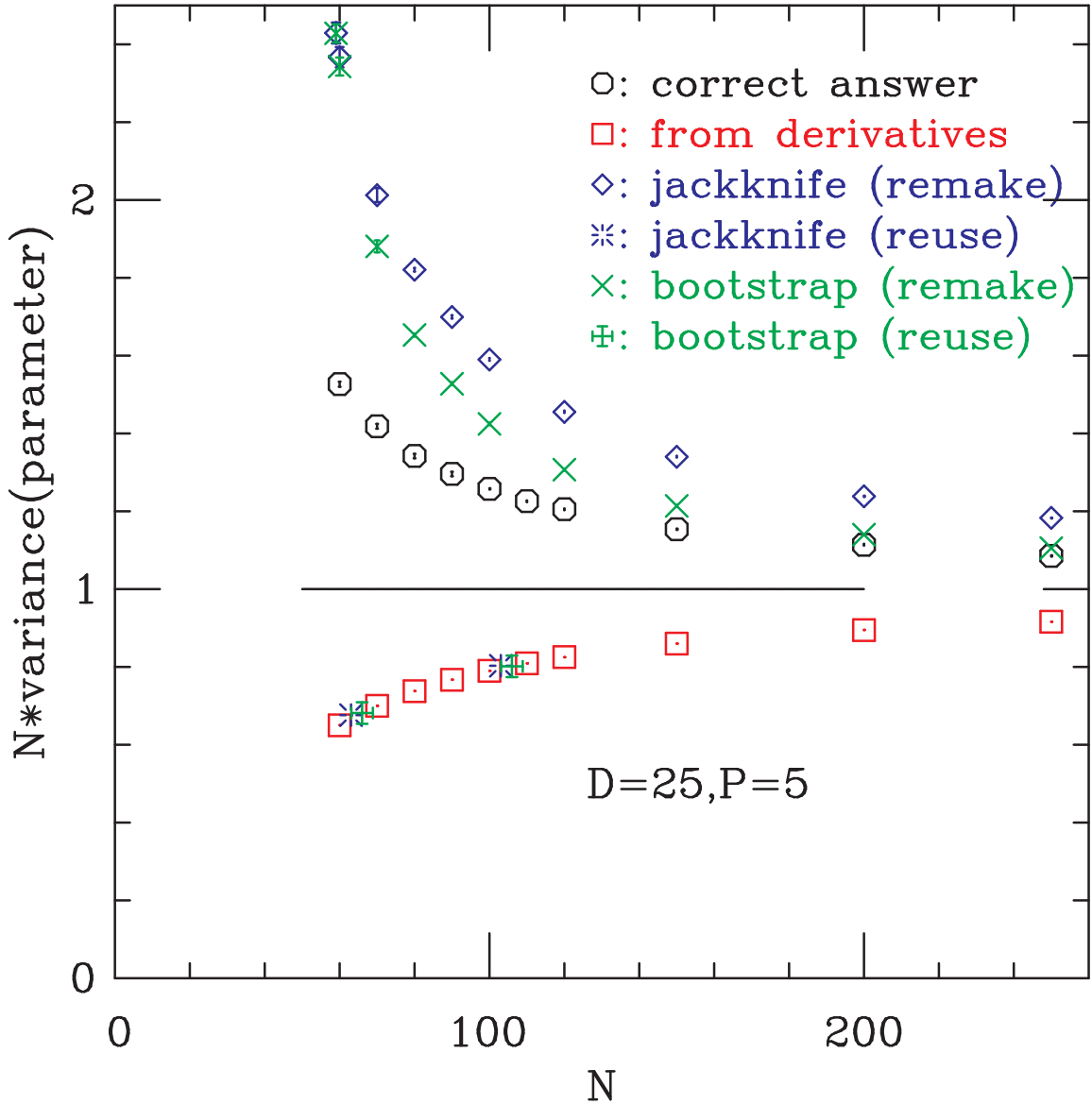}
\rule{0.0in}{0.0in}\vspace{1.0in}\\
\caption{
\label{FIG_sim1}
Number of samples times the variance (square of the error) of a parameter in fitting correlated data,
and averages over
trials of several methods for estimating this variance.
The horizontal line indicates the asymptotic value, $variance(x_0^f) = 1/N$.
$N$ is the number of samples; the meaning of
the plot symbols is described in the text.
}
\end{figure}

A common problem in analysis of experiments or of Monte Carlo
simulations is fitting a parameterized model to the average over
a number of samples of correlated data values.   In particular,
lattice QCD calculations typically require fitting operator
correlators, which are a function of distance between the
operators, to sums of exponentials with unknown amplitudes and
masses.
If the number of
samples is not infinite, estimates of the variance of the parameters
(``error bars'') and of the goodness of fit are affected.
This can be viewed as a generalization of the well known rule
``replace $N$ by $N-1$ in the denominator'' in calculating
the error on an average to the case where the error is on
a parameter estimated by a fit to correlated data points.
We calculate
approximate corrections to the variance of the parameters (see Fig.~\ref{FIG_sim1}
for a graphical example)
for estimates made in the standard way from derivatives
of the parameters' probability distribution as well as from jackknife
and bootstrap estimates.  (The distribution of parameter estimates is not
exactly Gaussian, so the variance of the parameters is not quite the
whole story.)
Without compensating for sample size effects, {\bf none} of these
methods give unbiased estimates of the parameters' variance.

Many numerical simulation programs or experiments involve two or more
stages of fitting, where the parameters resulting from the first stage
are the data input to the second stage.   For example, in computations
of meson decay constants in lattice QCD the first stage involves
fitting a correlator of meson operators to exponentials and extracting
the mass and amplitude, and the second stage involves fitting these
masses and amplitudes to functions of the quark masses and lattice
spacings to allow extrapolation to the chiral and continuum limits.
(See for example Refs.~\cite{lightlight} and \cite{heavylight}.)
For example, in Ref.~\cite{lightlight} about 600 hadron 
masses and amplitudes are computed in the first stage of fitting,
and these 600 numbers and their (co)variances are in turn the data
for the fitting in the second stage.
While an unbiased estimate of the variance of the parameters
is always welcome,
it is particularly important in this case since many parameters
in the first stage of fitting
are used as inputs (data) in the second stage of fitting, and if their
errors are systematically too large or too small the apparent
goodness of fit
in the second stage of fitting will be very good or very bad
respectively.

\section{The problem}
We consider a problem where we need to fit a function of $P$ parameters
to an average of $N$ samples, where each sample consists of $D$ data
points.   We use subscript indices to label the component of
the data vectors and superscript indices to label the samples.
Thus $x_i^a$ is the $i$'th component of the $a$'th sample, with
$0\le i<D$ and $0 \le a<N$.  Each sample is assumed to be normally distributed, but the
different components of the $D$ dimensional sample are generally
correlated.  Averages over samples will be denoted by
overbars. We will need to imagine
averaging over many trials of
the experiment, and we will use angle brackets to denote such
an average: $\big\langle \OL{x_i} \big\rangle$.

So, for example
\BEA
\la{EQ_oldef}
\OL{ x_i} &=& \OON\sum_a x_i^a \EL
\OL{x_i x_j} &=& \OON\sum_a x_i^a x_j^a\EL
\OL{x_i} \OL{x_j} &=& \OON\sum_a x_i^a  \OON \sum_b x_j^b\\
\EEA

The covariance matrix (``of the mean'') for one trial is
\BNE \label{EQ_covarmat_def}
C_{ij} = \OON \LP \OL{x_i x_j} -  \OL{x_i}\ \OL{x_j} \RP
  = \OOX{N^2} \sum_a x_i^a x_j^a - \OOX{N^3} \LP \sum_a x_i^a \RP \LP
\sum_b x_j^b \RP \ENE
This covariance matrix will fluctuate around the true covariance matrix,
obtainable only in the limit $N \rightarrow\infty$.
Note we use $\OON$ instead of $\OOX{N-1}$ in normalizing $C_{ij}$.
For our purposes, the difference
between these normalizations is best included with the other order $\OON$
effects to be discussed.

Fit parameters $p_\alpha$, with $0\le\alpha<P$, are obtained by minimizing
\BNE \chi^2 = \LP \OL{x_i} -x_i^f(p_\alpha) \RP \LP C^{-1} \RP _{ij}
\LP \OL{x_j} -x_j^f(p_\alpha) \RP \la{EQ_chisq_def}\ENE
where $x_i^f(p_\alpha)$ is the value of $\OL{x_i}$ predicted by the model.
As pointed out in Ref.~\cite{MICHAEL1}, since we are stuck with 
estimates of the covariance matrix and the $\OL{x_i}$
obtained from the same samples, they are correlated.

First change to a convenient coordinate system (alas, available
only in theory, not in practice).   For the moment we assume that
our fit model is good, so that the $x_i^f(p_\alpha)$ can be adjusted
to equal the true averages of the $x_i$.  Shift the coordinates so
that $\langle \OL{x_i} \rangle$ is zero.  Then rotate the coordinates
so that the {\bf true} covariance matrix is diagonal, and rescale
them so that  $\langle (x_i^a)^2 \rangle = 1$.  (So far, we have
followed Ref.~\cite{MICHAEL1}.)
We now have $\langle x_i^a  x_j^b \rangle = \delta_{ij} \delta^{ab} $,
and the true covariance matrix is the unit matrix.

Make a further rotation so that the changes in the $x_i^f(p_\alpha)$
as the $p_\alpha$ vary around their true values are in the first
$P$ components, and so that the changes in the $x_i^f(p_\alpha)$
as the first parameter $p_0$ varies are in the first component.
Now we can rescale $p_0$ so that $\PAR{p_0}{x_0}=1$, which simply
means that $p_0$ is the average $\OL{x_0}$.  In doing this we have
assumed that $p_0$ is linear enough in the $\OL{x_i}$ or that the
fluctuations in the $\OL{x_i}$ are small enough.

In this basis, write the covariance matrix (from the data in this experiment)
and its inverse in blocks,
\BEA \la{EQ_covmat_blocks}
C &\equiv& \LP \begin{array}{cc} U & V \\ V^T & W \end{array} \RP \EL
C^{-1} &\equiv& \LP \begin{array}{cc} A & B \\ B^T & E \end{array} \RP \\
\EEA
where the matrices $U$ and $A$ are $P$ by $P$, $V$ and $B$ are $P$ by $D-P$ and $W$ and $E$ are $D-P$ by $D-P$.

Now $\chi^2$ is given by
\BNE \chi^2 = \LP \OL{x_i} - x_i^{f} \RP \LP C^{-1} \RP _{ij} \LP \OL{x_j} - x_j^{f} \RP \label{chisq_def2}\ \ \ ,\ENE
where only the first $P$ components of $x_i^f$ are nonzero.
For example, with two parameters
\BNE \chi^2 = \LP \OL{x_1} - x_1^{f}, \OL{x_2} - x_2^{f}, \OL{x_3}, \ldots \RP
\LP \begin{array}{cc} A & B \\ B^T & E \end{array} \RP
\LP \begin{array}{c} \OL{x_1} - x_1^{f} \\ \OL{x_2} - x_2^{f} \\ \OL{x_3} \\ \ldots \end{array} \RP \ENE

The $x_i^f$ are found from minimizing $\chi^2$:
\BNE 0 = \frac{\partial\chi^2}{\partial x_\is^f}
= 2 A_{\is\js}\LP \OL{x_\js}-x_\js^f \RP + 2 B_{\is\jp} \OL{x_\jp} \ENE
where here and in many subsequent equations starred indices run from $0$ to $P-1$ and
primed indices from $P$ to $D-1$, and the factor of two comes from differentiating
with respect to the $x_i^f$ on both sides of Eq.~\ref{chisq_def2}
and using the fact that $C^{-1}$ is symmetric.

This is solved by
\BNE \label{EQ_xf_solution} x_\is^f = \OL{x_\is} + A_{\is \js}^{-1}B_{ \js \kp} \OL{x_\kp} \ENE
From $C\,C^{-1} = {\bf 1}$ and Eq.~\ref{EQ_covmat_blocks},
\BEA
\label{EQ_CCinv_eqs}
 UA+VB^T &=& {\bf 1} \EL
 UB+VE &=& {\bf 0} \EL
 V^TA+WB^T &=& {\bf 0} \EL
 V^TB+WE &=& {\bf 1} \\
\EEA
Using the third of Eqs.~\ref{EQ_CCinv_eqs}, remembering that $A$ and $W$ are symmetric, we get
an alternate to Eq.~\ref{EQ_xf_solution}:
\BEA \label{EQ_xf_solution_2}  B &=& -AVW^{-1}  \EL
     A^{-1}B &=& -VW^{-1} \EL
     x_\is^f &=& \OL{x_\is} - V_{\is\jp}W_{\jp\kp}^{-1}\OL{x_\kp} \\
\EEA
From this equation we see that, in this basis, parameter number zero, $x_0^f$, does
not depend on the other of the first $P$ components, $\OL{x_\is}$ with $1\le\is < P$.
Thus the distribution of parameters depends only
on the combination $D-P \equiv d$.

Similarly for $\chi^2$:
\BEA \chi^2 &=& \LP -\OL{x} B^T A^{-1}, \OL{x} \RP \LP \begin{array}{cc} A & B \\ B^T & E  \end{array} \RP
\LP \begin{array}{c} -A^{-1}B \OL{x} \\ \OL{x} \end{array} \RP \\
&=& \OL{x_\ip} \LP -B^TA^{-1}B+E \RP \OL{x_\jp}
\EEA
Now insert $W^{-1}W = {\bf 1}$ and use the third and fourth equations in \ref{EQ_CCinv_eqs}
\BEA \label{EQ_chisq_1}
   \chi^2 &=& \OL{x} W^{-1} \LP -WB^T A^{-1}B+WE \RP \OL{x} \EL
    &=& \OL{x} W^{-1} \LP V^T A A^{-1}B+WE \RP \OL{x} \EL
    &=& \OL{x} W^{-1} \LP V^T B+WE \RP \OL{x} \EL
    &=& \OL{x_\ip} W_{\ip\jp}^{-1} \OL{x_\jp}  \\
\EEA
But $W$ is just the covariance matrix for the last $D-P$ components of $\OL{x}$ in this basis,
so the statistical properties of $\chi^2$ are exactly the same as a $D-P$ dimensional problem
with no fit parameters, and the distribution of $\chi^2$, as expected, depends
only on the number of degrees of freedom, $\dof \equiv D-P$.
We note that the distribution of $\chi^2$ (more properly, $T^2$) is known.
Since it is important here and closely related to the estimates of
parameter errors, we quote the result in Appendix I.

\section{Numerical example}

To illustrate the effects of sample size, we begin with a numerical example,
using the basis described above.
In this example, $N$ Gaussian distributed random data vectors  with $D=25$ were generated.
The data was fit with $P=5$ parameters, which are just the first $P$ components
of the average data vector. This was repeated for many trials.
The black octagons in Fig.~\ref{FIG_sim1} show $N$ times the
variance (over trials) of one of the parameters, where the asymptotic value is one.
These black octagons are the correct answer for the variance of the parameter, and this is
the variance that we wish to estimate from our experiment, where
we only have one trial to work with.  We see that for finite $N$ the parameters
fluctuate by an amount larger than the asymptotic value.

We also show the average over trials of the variance estimated from derivatives
of the parameter probability, the average over trials of the variance
estimated from a single elimination jackknife analysis, and the average
from a bootstrap analysis.  For the jackknife and bootstrap, the plot contains
average variances both for the case where the full sample covariance matrix
was used in each resampling
and where a new covariance matrix was made
using the data in each jackknife or bootstrap sample.
Red squares are average variances from the
usual ``derivative'' method.  Blue diamonds are from a single elimination
jackknife analysis where a new covariance matrix was made for each
jackknife sample. The two blue bursts (on top of
the red squares) use the full sample covariance matrix in each jackknife resample.
Similarly, the green fancy plusses are from a bootstrap
analysis, using the covariance matrix from the original sample.
The green crosses are estimates from a bootstrap analysis where a new
covariance matrix was made for each bootstrap sample.

We see that correct answer deviates from the asymptotic value
for finite $N$, and that the various methods for estimating this variance
produce biased estimates of the variance of the parameter.

\section{Large $N$ expansion}

Most of the effects shown in Fig.~\ref{FIG_sim1} can be understood
analytically.
We can expand the covariance matrix in each trial around its true
value,
\BNE C_{ij} = \OON \LB \delta_{ij} + \LP \OL{x_i x_j} - \delta_{ij} -
\OL{x_i}\OL{x_j} \RP \RB \ENE
Here the term in parentheses has fluctuations of order $1/\sqrt{N}$
and an average of order $\OON$.
Thus its square will also have expectation value  $\approx \OON$.

Then
\BEA C_{ij}^{-1} &=& N \delta_{ij} \EL
 &-& N \LP \OL{x_i x_j} - \delta_{ij} - \OL{x_i}\OL{x_j} \RP \EL
&+& N \LP \OL{x_i x_k} - \delta_{ik} - \OL{x_i}\OL{x_k} \RP
\LP \OL{x_k x_j} - \delta_{kj} - \OL{x_k}\OL{x_j} \RP \EL
&+& \ldots \\
\EEA

Using the fact that integrals of polynomials weighted by Gaussians are found
by pairing the $x_i^a$ in all possible ways, or making all possible
contractions, we can develop rules for calculating these expectation
values.  We will use parentheses to list the pairings.  For example,
with $(12)$ indicating that the first and second $x$ are paired,
\BEA
&& \LL \OL{x_i} \OL{x_i x_j} \OL{x_j}\RR \EL
&=& \OL{x_i} \OL{x_i x_j} \OL{x_j} (12)(34) \EL
&+&  \OL{x_i} \OL{x_i x_j} \OL{x_j} (13)(24) \EL
&+& \OL{x_i} \OL{x_i x_j} \OL{x_j} (14)(23) \\ \EEA

Using 
\BE\label{EQ_basic_delta}\langle x_i^a  x_j^b \rangle = \delta_{ij} \delta^{ab} \EE
and
\BE \label{EQ_basic_bar}\OL{x_i}= \OON\sum_a x_i^a \EE

we get the Feynman rules for contractions of barred quantities.
\begin{enumerate}

\item{Each contraction gives a $\delta_{ij}$ for the lower indices it connects.}

\item{Each bar gives a $\OON$, whether it covers a single $x$ or two, $\OL{x_i}$ or
$\OL{x_i x_j}$ --- see Eq.\ref{EQ_oldef}. }

\item{Each continuous line made of overbars and contraction symbols gives a factor
of $N$.  This is from the $\sum_{ab\ldots} \delta^{ab}\delta^{bc}\ldots$, which has
$N$ nonzero terms. For example, $\OL{x_i}\OL{x_j x_k}\OL{x_l} (12)(34)$ is one continuous
line, while $\OL{x_i}\OL{x_j x_k}\OL{x_l} (14)(23)$ is two lines (one is a loop).  This
results in every loop giving an extra factor of $N$ relative to other contractions with
the same number of fields. }

\end{enumerate}

Since an open line (not a loop) with $C$ contractions has $2C$ $x$'s and $C+1$ bars,
but a loop with $N$ contractions has $2C$ $x$'s and $C$ bars, these rules can be rephrased as:
\begin{enumerate}
\item{Each contraction gives a $\delta_{ij}$ for the lower indices it connects.}
\item{Each $x$ gives a factor of $1/\sqrt{N}$.}
\item{Each loop gives a factor of $N$.}
\end{enumerate}

In the expansion of $C^{-1}$ we find the combination
$\OL{x_i x_j} - \delta_{ij} - \OL{x_i}\OL{x_j}$, which we will
denote by $\OLL{x_i x_j}$.  This occurs frequently enough that we should
state special rules for it.

In evaluating an expression  containing $\OLL{x_i x_j}$ there will be contractions
where the $x_i$ and $x_j$ in $\OL{x_i x_j}$ are contracted with each other.
 These contractions just cancel the $\delta_{ij}$.  The terms with $\OL{x_i}$
and $\OL{x_j}$ contracted give a $-\delta_{ij}/N$.
Thus a ``tadpole'' where $\OLL{x_i x_j}(12)$ contracts with itself just gives a
$-\delta_{ij}/N$. (This includes the $N^{-1/2}$ from each of the $\OL{x}$'s.)

Now consider terms where $\OLL{x_i x_j}$ is part of an open line, like
\BNE \OL{x_i} \OLL{x_i x_j} \OL{x_j} (12)(34) \ENE
In this case the $\OL{x_i x_j}$ and the $-\OL{x_i}\OL{x_j}$ cancel,
so $\OLL{x_i x_j}$ can never be part of an open line.
But if this object is part of a loop, like in
\BNE \OLL{x_i x_j} \OLL{x_k x_l} (13)(24) \ENE
the $\OL{x_i x_j}$ part is part of the loop, but the $-\OL{x_i}\OL{x_j}$ part
breaks the loop.  Thus the four paths hidden in these double bars give
\BNE N-2+1 = (N-1)\delta_{ik} \delta_{jl} \ENE
Similarly, a loop of three double bars gives $2^3=8$ terms, $N-3+3-1 = (N-1)$,
and any loop made up entirely of $\OLL{x_i x_j}$'s gives a factor of $N-1$ times
the appropriate Kronecker $\delta$'s.

As trivial examples,
\BEA \LL \OL{x_i} \OL{x_j} \RR &=& \OL{x_i} \OL{x_j} (12) = \OON\delta_{ij} \\
 \LL \OL{x_i x_j} \RR &=& \OL{x_i x_j} (12) = \delta_{ij} \\
 \LL N\, C_{ij} \RR &=& 1+\OLL{x_i x_j}(12) = \LP 1 - \OON \RP \delta_{ij} \\
\EEA

We are also interested in the variances of averaged quantities.
For the variance of something, $var(X) = \LL X^2 \RR - \LL X \RR^2$,
we need the ``connected part'' of $\LL X^2 \RR$.  We use a vertical bar to denote
this, and we only need contractions where some of the lines cross the bar.

As an example, for the variance of an arbitrary element of $C$ to lowest order,
\BEA \label{EQ_connpart_1}  \LL var\LP C_{ij} \RP \RR
         &=& \LL C_{ij}^2 \RR - \LL C_{ij} \RR_{no\ sum\ ij}^2 \\
&=& \OOX{N^2} \LL 1+\OLL{x_i x_j} \Big|\  1+\OL{x_i x_j} \RR_{NS}\la{varc_2_eq}\\
&=& \OOX{N^2} \LP \OLL{x_i x_j} \Big|\  \OLL{x_i x_j} \RP (13)(24)+(14)(23) \EL
&=& \frac{N-1}{N^4} \LP \delta_{ii}\delta_{jj} + \delta_{ij}\delta_{ij} \RP_{NS} \EL
&=& \OOX{N^2} \frac{2}{N} ;\ i=j \EL
&& \OOX{N^2} \frac{1}{N} ;\ i\ne j \\
\EEA
The last line is written to display that the fractional variance on
the diagonal element is $\frac{2}{N}$.

In equations where the components are separated into starred indices,
$0 \le \is <P$ and primed indices, $P \le \ip < D$,
contractions of primed with starred indices are zero, contractions of starred with
starred indices give delta functions with $\delta_{\is\is} = P$, and primed with primed use
$\delta_{\ip\ip} = D-P$.

\section{Variance (and higher moments) of the parameters}

In this section we examine the variances of the parameters --
that is, the error bars on our answers.  First we calculate how
much the parameters actually vary over many trials of the experiment.
Then we calculate the average of common ways of estimating this
variance --- from derivatives of the probability, from
an ``eliminate J'' jackknife analysis
or from a bootstrap resampling
(using either the covariance
matrix from the full sample, or a new covariance matrix made from
each jackknife or bootstrap resample).
The differences allow us to find and correct for
bias in our error estimates resulting from the finite sample size.

For the actual variance of our parameters, use Eq.~\ref{EQ_xf_solution_2}.
Since we are in a coordinate system where the average of this quantity is
zero, we don't need to worry about taking the connected part.

\BEA \label{EQ_par_var_1}
\LL x_0^f x_0^f \RR &=& \LL \OL{x_0}\OL{x_0}  \RR \label{varreal_p0_eq} \EL
 &+& 2\LL \OL{x_0} V_{0 \jp} W_{ \jp \kp}^{-1} \OL{x_\kp} \RR \\
 &+& \LL \OL{x_\jp} W_{\jp \kp }^{-1} V_{\kp 0}^T V_{0 \mp } W_{\mp \np}^{-1} \OL{x_\np} \RR
\EEA
Since $V$ and $W$ are made entirely of double bars and can therefore only be part
of a loop, and primed indices can't contract with index zero,
the middle term (cross term) is zero.  

We compute this to order $\frac{1}{N^3}$.
\BEA
\LL x_0^f x_0^f \RR &=&  \LL \OL{x_0}\OL{x_0}  \RR \EL
 &+& \Big\langle \OL{x_\jp} \LP \delta_{\jp \mp} - \OLL{x_\jp x_\mp} + \OLL{x_\jp x_\pp} \OLL{x_\pp
x_\mp} \ldots \RP 
\LP \OLL{x_\mp x_0} \RP  \EL
&& \LP \OLL {x_0  x_\np} \RP
\LP \delta_{ \np \kp} - \OLL{x_\np x_\kp} + \OLL{x_\np x_\rp}\OLL{x_\rp x_\kp} \ldots \RP \OL{x_\kp} \Big\rangle \\
\EEA

The leading term, $\LL \OL{x_0}\OL{x_0}  \RR$, is just $\OON$.

The term with six $x$'s has only one contraction:
\BEA && \OL{x_\jp} \OLL{x_\jp x_0} \OLL{x_0  x_\kp} \OL{x_\kp} \ \ (16)(25)(34) \EL
&=& \frac{N-1}{N^3} d \\ \EEA
where $d \equiv D-P$.

There are two equal terms with eight $x$'s.
There are three nonzero contractions of this term.  Ignoring the $N^{-4}$ parts,
these are
\BEA
- 2\, \OL{x_\jp} \OLL{x_\jp x_\mp} \OLL{x_\mp x_0} \OLL{x_0 x_\kp} \OL{x_\kp} 
\ (18)(27)(34)(56) &=& \frac{-2}{N^3} \delta_{\jp\kp}\delta_{\jp\kp} \delta_{\mp\mp} \delta_{00} \EL
- 2\, \OL{x_\jp} \OLL{x_\jp x_\mp} \OLL{x_\mp x_0} \OLL{x_0 x_\kp} \OL{x_\kp} 
\ (18)(23)(47)(56) &=& \frac{+2}{N^3} \delta_{ \jp \kp}\delta_{ \jp \mp } \delta_{ \mp \kp} \delta_{00} \EL
- 2\, \OL{x_\jp} \OLL{x_\jp x_\mp} \OLL{x_\mp x_0} \OLL{x_0 x_\kp} \OL{x_\kp} 
\ (18)(24)(37)(56) &=& \frac{-2}{N^3} \delta_{ \jp \kp}\delta_{ \jp \mp } \delta_{ \mp \kp} \delta_{00} \\
\EEA
Here the plus sign on the second contraction comes from the tadpole.  The
second and third contractions cancel, so we just have
\BNE \frac{-2}{N^3} d^2 \ENE

To order $\OOX{N^3}$ we only need two loop contractions from the terms with ten $x$'s.
There are three such terms, but two of them are equal.
\BEA && \LL \OL{x_\jp} \OLL{x_\jp x_\mp} \OLL{x_\mp x_0} \OLL{x_0 x_\np} \OLL{ x_\np x_\kp} \OL{x_\kp} \RR \EL
+2 && \LL \OL{x_\jp} \OLL{x_\jp x_\pp} \OLL{x_\pp x_\mp} \OLL{x_\mp x_0} \OLL{x_0 x_\kp} \OL{x_\kp} \RR \\ \EEA
Each term has two contractions:
\BEA
&& \OL{x_\jp} \OLL{x_\jp x_\mp} \OLL{x_\mp x_0} \OLL{x_0 x_\np} \OLL{x_\np x_\kp} \OL{x_\kp}
\ (1,10)(28)(39)(47)(56) \EL
&+&  \OL{x_\jp} \OLL{x_\jp x_\mp} \OLL{x_\mp x_0} \OLL{x_0 x_\np} \OLL{ x_\np x_\kp} \OL{x_\kp}
\ (1,10)(29)(38)(47)(56) \EL
&=& \OOX{N^3} \LP d + d^2 \RP \\ \EEA
\BEA
&& 2\  \OL{x_\jp} \OLL{x_\jp x_\pp} \OLL{x_\pp x_\mp} \OLL{x_\mp x_0} \OLL{x_0 x_\kp} \OL{x_\kp}
\ (1,10)(24)(35)(69)(78) \EL
&+& 2\
 \OL{x_\jp} \OLL{x_\jp x_\pp} \OLL{x_\pp x_\mp} \OLL{x_\mp x_0} \OLL{x_0 x_\kp} \OL{x_\kp}
\ (1,10)(25)(34)(69)(78) \EL
&=& \frac{2}{N^3} \LP d + d^2 \RP \\ \EEA

Putting it all together,
\BEA \label{EQ_par_var_2} \LL x_0^f x_0^f \RR &=& \OOX{N} + \frac{N-1}{N^3} (d)
+ \frac{-2}{N^3} (d)^2 + \frac{3}{N^3} (d+d^2) \EL
&=& \OOX{N} + \frac{d}{N^2} + \frac{d(d+2)}{N^3} + \ldots \\ \EEA
Thus the fluctuations in the parameters are larger than the asymptotic value $\OON$.
from the covariance matrix.

As noted above, the probability distribution of the parameters is not exactly
Gaussian.   Higher moments of this distribution can be obtained in the same
way.   At leading order in $\OON$ there is only one independent diagram for
the connected part of each moment, and we find, for $M$ even,
\BNE \LL \LP x_0^f \RP^M \RR_{connected} = \frac{ \LP D-P \RP \LP M-1 \RP ! }{N^M} \ENE

\section{Estimates of the parameters' variance}

In practice, the most common method for estimating the variance of
the parameters is to use the covariance matrix for the parameters.
(See, for example, Ref.~\cite{TASI}.)
In our coordinate system, this matrix is just $A^{-1}$, and our estimate
for the variance of parameter zero is $\LP A^{-1}\RP_{00}$.
Using the third and first of Eqs.~\ref{EQ_CCinv_eqs},
\BEA
B^T &=& - W^{-1} V^T A \EL
{\bf 1} &=& UA-VW^{-1}V^TA \EL
A^{-1} &=& U - VW^{-1} V^T \\
\EEA
Then, our estimate for the variance of parameter zero is
\BEA
 var(x_0^f)_{derivative} &=&  A_{00}^{-1} \EL
 &=& U_{00} - V_{0\kp} W_{\kp \lp}^{-1} V_{\lp 0}^T \EL
&=& \OON \LP \delta_{00} + \OLL{x_0 x_0} \RP \EL
&-& \OON \LP \OLL{x_0 x_\kp} \RP \LP \delta_{\kp \lp } - \OLL{x_\kp x_\lp}
  + \OLL{x_\kp x_\mp} \OLL{x_\mp x_\lp} \ldots \RP \OLL{x_\lp x_{0}} \\
\EEA
For the order $\OON$ correction we only need the $\delta_{\kp \lp }$ from
$W^{-1}$, and find

\BEA
\label{EQ_varest_p0}
 N var(x_0^f)_{derivative}
&=& \delta_{00} 
 + \OLL{x_0 x_0}(12) 
 - \OLL{x_0 x_\kp} \OLL{ x_\kp x_0} (14)(23) \EL
 &=& 1 
 - \OON 
 - \frac{N-1}{N^2}(D-P) \EL
&=& 1 - \OON\LP 1+D-P\RP \\
\EEA
The order $1/N^2$ contribution to this estimate vanishes, as sketched in
Appendix II.
If $D=P=1$ this is just $ \OON \LL 1+\OLL{x_0 x_0}\RR = 1 - \OON $,
the standard correction for a simple average, reflecting our normalization
of the covariance matrix.  Comparing to the desired result in Eq.~\ref{EQ_par_var_2},
we see that this is an underestimate of the variance of the parameters.
The difference between this error estimate and the correct one above
is that this estimate assumes that the covariance matrix remains
fixed while the data points vary, while the correct answer takes into account
the correlations between the data points and the covariance matrix
(constructed from these same data points).


\section{Variance of jackknife and bootstrap parameters}

The variance of the parameters is also often estimated by a jackknife
or bootstrap analysis.  In these methods the fit is repeated many times
using subsets of the data sample, and the variance of the parameters
is estimated from the variance over the jackknife or bootstrap samples.
Both the jackknife and bootstrap can
be done either using the covariance matrix from the full sample in fitting
each jackknife or bootstrap sample, or by remaking a covariance matrix for
each resample.   Using the full sample covariance matrix amounts
to seeing how the parameters vary with fixed covariance matrix, that is,
by varying $ \OL{x_\is}$ and $\OL{x_\kp}$ in Eq.~\ref{EQ_xf_solution}
with $A_{\is \js}^{-1}B_{ \js \kp}$ held fixed.
This is
the same question as is answered by $var(x_0^f)_{derivative}$ in Eq.~\ref{EQ_varest_p0}.
Since the change in the parameters is linear in $ \OL{x_\is}$ and $\OL{x_\kp}$,
it doesn't matter if the $\OL{x_i}$  are varied infinitesimally (by taking derivatives)
or slightly (jackknife) or fully (bootstrap).
In this case, the variance of the parameters will have the
same bias as does Eq.~\ref{EQ_varest_p0} --- no new calculation
is necessary, although there is a slight difference due to the
normalization of the covariance matrix used here.

Remaking the covariance matrix for each resample includes correlations
of the covariance matrix and data, but not in quite the desired way.
The calculations above can be extended to calculate the expectation value
of the parameter variance for the jackknife analysis in which the covariance
matrix is recomputed for each jackknife sample.
An ``eliminate J'' jackknife consists of making $N/J$
resamples, each omitting $J$ data vectors (numbers $nJ$ through $(n+1)J-1$),
and hence having $N_J \equiv N-J$ elements.
We will denote averages in the $n$'th jackknife sample with a superscript
$(n)$.
The average of $x^a$ in the $n$'th jackknife sample is
\BNE \OL{x^{a(n)}} = \frac{1}{N_J} \LP \sum_{a \in (n)}x^a \RP \ENE
where $J$ data vectors (starting with number $nJ$) were deleted from the full
sample.  The variance of this quantity (over the jackknife samples) is
\BNE\label{EQ_jack_var_2} \frac{J}{N(N-J)} \ \ \ ,\ENE
so we generally multiply the variance over the jackknife samples
by $\frac{N-J}{J}$ to get the expected variance of the mean $\OON$.

We now compute the variance of the parameters in the jackknife fits.  In doing
this we will need averages of products of quantities from different
jackknife ensembles.  Without losing generality, we can think of these
as ensembles number zero and one, which differ only in their first
$J$ data elements.  Thus, expectation values of sums over values in
different ensembles may produce factors of $N_J-J$ instead of
$N_J$, where $N_J$ is the number of samples in the jackknife,
and is really $N-J$. ($N_J-J = N-2J$ is the number of samples in common between
two different jackknife resamples.)

For example, using $(n)$ to denote quantities in the $n$'th jackknife
sample ($x_j^{a(n)}$ is the $j$'th component of the $a$'th data vector in
jackknife sample $(n)$), for $n \ne m$,
\BEA \label{EQ_jack_avg1} && \langle \OL{x_{j}^{(n)}} \OL{x_{k}^{(m)}} \rangle \EL
 &=& \langle \frac{1}{N_J} \sum_a x_j^{a(n)} \frac{1}{N_J} \sum_b x_k^{b(m)} \rangle \EL
 &=& \frac{1}{N_J^2} \sum_{ab} \delta_{jk} \deltabar^{ab} \EL
 &=& \frac{N_J-J}{N_J^2} \delta_{jk} \\
\EEA
where we define $\deltabar^{ab} = 1$ if $a=b$ and $a,b \in (J,N-1)$, $0$ otherwise.   Thus the
sum over $a$ and $b$ gives a factor of $N_J-J$ instead of $N_J$.

From Eq.~\ref{EQ_xf_solution_2}, parameter $0$ in jackknife fit $(n)$ is
\BNE x_0^{(n)f} = \OL{x_0^{(n)}} + V_{0\jp}^{(n)} W_{\jp\kp}^{-1(n)} \OL{x_\kp^{(n)}} \ENE
and the variance of this parameter over the jackknife samples is
\BEA\label{EQ_jack_var}  var_J(x_0^f) &=&
\Bigg\langle \LP \OL{x_0^{(n)}}  - \OL{x_\jp^{(n)}} W_{\jp \ip}^{-1(n)} V_{\ip 0}^{T(n)}
- \frac{J}{N} \sum_m \LP \OL{x_0^{(m)}} - \OL{x_\kp^{(m)}} W_{\kp \ip}^{-1m)} V_{\ip 0}^{T(m)} \RP \RP
 \EL
&& \LP \OL{x_0^{(n)}} - V_{0\jp}^{(n)} W_{\jp\kp}^{-1(n)} \OL{x_\kp^{(n)}}
- \frac{J}{N} \sum_p \LP \OL{x_0^{(p)}} -  V_{0\jp}^{(p)} W_{\jp\lp}^{-1(p)} \OL{x_\lp^{(p)}} \RP \RP
 \Bigg\rangle \\
\EEA

This is more complicated than Eq.~\ref{EQ_par_var_1} because the mean
over jackknife samples is not exactly zero.  Also, the sums over sample vectors now
sometimes give $N_J$, sometimes $N_J-1$, and sometimes $N_J-J$, so some of the shortcuts developed
above won't work any more.  Note the $J/N$ is correct -- there are $N/J$ jackknife
resamples, each containing $N_J = N-J$ elements.

In Eq.~\ref{EQ_jack_var} the sums contain terms where $n=m$ and terms where $n \ne m$.
Separate the diagonal and off-diagonal terms in the sums, and use the fact that
all non-diagonal terms are equal, $\sum_m$ contains $N/J-1$ terms with $m \ne n$, 
and $\sum_{mp}$ has $N/J$ diagonal terms and $(N/J)(N/J-1)$ off diagonal:
\BEA var_J(x_0^f) = \Bigg \langle  &&
    \LP 1-\frac{J}{N} \RP \OL{x_0^{(n)}} \OL{x_0^{(n)}} \EL
 &-& \LP 1-\frac{J}{N} \RP \OL{x_0^{(n)}} \OL{x_0^{(m)}} \EL
 &-& 2 \LP 1-\frac{J}{N} \RP \OL{x_0^{(n)}}
    V_{0\ip}^{(n)} W_{\ip\kp}^{-1(n)} \OL{x_\kp^{(n)}} \EL
 &+& 2 \LP 1-\frac{J}{N} \RP \OL{x_0^{(n)}} 
   V_{0\ip}^{(m)} W_{\ip\kp}^{-1(m)} \OL{x_\kp^{(m)}} \EL
 &+& \LP 1-\frac{J}{N} \RP \OL{x_\jp^{(n)}} W_{\jp \ip}^{-1(n)} V_{\ip 0}^{T(n)}
    V_{0\ip}^{(n)} W_{\ip\kp}^{-1(n)} \OL{x_\kp^{(n)}} \EL
 &-& \LP 1-\frac{J}{N} \RP \OL{x_\jp^{(n)}} W_{\jp \ip}^{-1(n)} V_{\ip 0}^{T(n)}
   V_{0\ip}^{(m)} W_{\ip\kp}^{-1(m)} \OL{x_\kp^{(m)}}
\Bigg\rangle_{n \ne m} \\
\EEA
where $n \ne m$.
To evaluate this expression we need:
\BEA \label{jackvar_parts_eq0}
(a)&&\ \ \langle \OL{x_0^{(n)}} \OL{x_0^{(n)}} \rangle \EL
(b)&&\ \ \langle \OL{x_0^{(n)}} \OL{x_0^{(m)}} \rangle_{n \ne m} \EL
(c)&&\ \ \langle \OL{x_0^{(n)}}  V_{0\ip}^{(n)} W_{\ip\kp}^{-1(n)} \OL{x_\kp^{(n)}} \rangle \EL
(d)&&\ \ \langle \OL{x_0^{(n)}} V_{0\ip}^{(m)} W_{\ip\kp}^{-1(m)} \OL{x_\kp^{(m)}} \rangle_{n \ne m} \EL
(e)&&\ \ \langle \OL{x_\jp^{(n)}} W_{\jp \ip}^{-1(n)} V_{\ip 0}^{T(n)} V_{0\jp}^{(n)} W_{\jp\kp}^{-1(n)}
   \OL{x_\kp^{(n)}} \rangle \EL
(f)&&\ \ \langle \OL{x_\jp^{(n)}} W_{\jp \ip}^{-1(n)} V_{\ip 0}^{T(n)} V_{0\jp}^{(m)} W_{\jp\kp}^{-1(m)}
   \OL{x_\kp^{(m)}} \rangle_{n \ne m} \\
\EEA
Here $(a)$, $(c)$ and $(e)$, which involve only jackknife sample $(n)$,
are the same as in the previous section with the
replacement of $N$ by $N_J$.
Because $W^{-1(n)}$ and $V^{(n)}$ consist only of double barred quantities which can't
be part of an open line, and primed and unprimed indices can't contract,
$(c)$ vanishes.
For $(d)$ we can imagine expanding all the $\OLL{x_i x_j}^{(m)}$'s into
pieces, $\LP \OL{x_i x_j}^{(m)} - \delta_{ij} - \OL{x_i}^{(m)} \OL{x_j}^{(m)} \RP$.
and making all contractions.  There is only one factor of $x$ from jackknife
sample $(n)$, which must contract with something from $(m)$.  Thus, all of
these terms differ from $(c)$ by replacement of exactly one factor of $N_J$ by $N_J-J$, and
therefore also sum to zero.

Similarly $(e)$ and $(f)$ differ by the replacement of one or more factors
of $N_J$ by $N_J-J$.  Thus, their difference will be one order in $\frac{J}{N}$
less than their value. 
This means that to get the first correction to the asymptotic form, we
need only keep the lowest order term in part $(e)$, and the analogous
contraction for part $(f)$.

\BEA \label{jackvar_parts_eq}
(a)&&\ \ \langle \OL{x_0^{(n)}} \OL{x_0^{(n)}} \rangle =
    \frac{1}{N_J^2} \sum_{ab} x_0^{a(n)} x_0^{b(n)} = \frac{1}{N_J} \EL
(b)&&\ \ \langle \OL{x_0^{(n)}} \OL{x_0^{(m)}} \rangle =
    \frac{1}{N_J^2} \sum_{ab} x_0^{a(n)} x_0^{b(m)} = \frac{N_J-J}{N_J^2} \EL
(e)&&\ \ \langle \OL{x_\jp^{(n)}} \OLL{x_\jp x_0^{(n)}} \OLL{x_0 x_\kp^{(n)}}
   \OL{x_\kp^{(n)}} \rangle \EL
   &&=\OL{x_\jp^{(n)}} \OLL{x_\jp x_0^{(n)}} \OLL{x_0 x_\kp^{(n)}}
   \OL{x_\kp^{(n)}} \ \ (16)(25)(34) \EL
   &&= \frac{N_J-1}{N_J^3}\LP D-P \RP \EL
(f)&&\ \ \langle \OL{x_\jp^{(n)}} \OLL{x_\jp x_0^{(n)}} \OLL{x_0 x_\kp^{(m)}}
   \OL{x_\kp^{(m)}} \rangle \EL
  &&= \OL{x_\jp^{(n)}} \OLL{x_\jp x_0^{(n)}} \OLL{x_0 x_\kp^{(m)}}
   \OL{x_\kp^{(m)}}  \ \ (16)(25)(34) \EL
  &&= \frac{N_J-2J-1 + \ldots}{N_J^3}\LP D-P \RP \\
\EEA
Here $(a)$, $(c)$ and $(e)$ are the same as in the previous section.
To evaluate $(f)$ we need to separate the two terms in the double overbar (the delta function
isn't there since the indices can never be equal), since they may give different
numbers of factors of $N_J-J$.  This evaluation proceeds as:

\BEA
(f) &\approx& \big\langle \OL{x_\jp^{(n)}} \OLL{x_\jp x_0^{(n)}} \OLL{x_0 x_\kp^{(m)}}
   \OL{x_\kp^{(m)}} \big\rangle_{n \ne m}  \EL
&=& \Big\langle \OL{x_\jp^{(n)}} \OL{x_\jp x_0^{(n)}} \OL{x_0 x_\kp^{(m)}}
   \OL{x_\kp^{(m)}}  \EL
&-& \OL{x_\jp^{(n)}} \OL{x_\jp^{(n)}} \OL{x_0^{(n)}} \OL{x_0 x_\kp^{(m)}}
   \OL{x_\kp^{(m)}}  \EL
&-&  \OL{x_\jp^{(n)}} \OL{x_\jp x_0^{(n)}} \OL{x_0^{(m)}} \OL{ x_\kp^{(m)}}
   \OL{x_\kp^{(m)}}  \EL
&+&  \OL{x_\jp^{(n)}} \OL{x_\jp^{(n)}} \OL{ x_0^{(n)}} \OL{x_0^{(m)}} \OL{ x_\kp^{(m)}}
   \OL{x_\kp^{(m)}} \Big\rangle_{n \ne m}  \EL
=\frac{1}{N_J^6} \sum_{abcdef} \Big\langle
&& x_\jp^{a(n)} x_\jp^{b(n)} x_0^{b(n)} x_0^{d(m)} x_\kp^{d(m)} x_\kp^{f(m)} \EL
&-& x_\jp^{a(n)} x_\jp^{b(n)} x_0^{b(n)} x_0^{d(m)} x_\kp^{e(m)} x_\kp^{f(m)} \EL
&-& x_\jp^{a(n)} x_\jp^{b(n)} x_0^{c(n)} x_0^{d(m)} x_\kp^{d(m)} x_\kp^{f(m)} \EL
&+& x_\jp^{a(n)} x_\jp^{b(n)} x_0^{c(n)} x_0^{d(m)} x_\kp^{e(m)} x_\kp^{f(m)} \Big\rangle_{n \ne m} \\
\EEA
Now the $(16)(25)(34)$ contraction gives
\BEA
=\frac{1}{N_J^6} \sum_{abcdef} \Bigg(
 && \delta_{\jp\kp} \deltabar^{af} \delta_{\jp\kp} \deltabar^{bd} \delta_{00} \deltabar^{bd} \EL
&-& \delta_{\jp\kp} \deltabar^{af} \delta_{\jp\kp} \deltabar^{be} \delta_{00} \deltabar^{bd} \EL
&-& \delta_{\jp\kp} \deltabar^{af} \delta_{\jp\kp} \deltabar^{bd} \delta_{00} \deltabar^{cd} \EL
&+& \delta_{\jp\kp} \deltabar^{af} \delta_{\jp\kp} \deltabar^{be} \delta_{00} \deltabar^{cd} \Bigg) \EL
= \frac{1}{N_J^6} && \Big(  N_J^2 \LP N_J-J \RP^2 -2 N_J \LP N_J-J \RP^2 + \LP N_J-J \RP^3 \Big) \LP D-P \RP \EL
= \frac{1}{N_J^6} && \Big(  N_J^4 -2N_J^3 J -N_J^3  + \ldots \Big) \LP D-P \RP \EL
= \frac{1}{N_J^3} && \Big( N_J-2J-1 + \ldots  \Big) \LP D-P \RP \\
\EEA

Putting the pieces together, the variance of the parameter over the jackknife
samples is
\BEA &&  \LP 1-\frac{J}{N} \RP \LP \frac{J}{N_J^2}  +  \frac{2J(D-P)}{N_J^3} \RP \\
 &=&  \LP \frac{N-J}{N} \RP \LP \frac{J}{(N-J)^2} \RP  \LP 1  +  \frac{2(D-P)}{N}  + \ldots \RP \\
\EEA

Comparing with Eq.~\ref{EQ_jack_var_2}, which is for $D=P$, we see that there
is an extra factor of $1 + \frac{2(D-P)}{N}$ (independent of $J$).
However, by comparison with Eq.~\ref{EQ_par_var_2}
we see that this effect is too large by a factor of two, so the jackknife
variance for the parameters is also biased.

The leading corrections to the bootstrap estimate of the parameters' variance
can be done in a similar way.
To be specific, our bootstrap procedure is to make $B$ resamplings, each made by
choosing $N$ data vectors with replacement from the original set of $N$ vectors,
and calculate the variance of the parameters over the bootstrap resamples.
Similarly to Eq.~\ref{EQ_jack_var}, the average over trials of the bootstrap estimate of the 
variance is
\BEA\label{EQ_bootvar1} var_B(x_0^f) &=&
\Bigg\langle  \LP \OL{x_0^{(n)}}  - \OL{x_\jp^{(n)}} W_{\jp \ip}^{-1(n)} V_{\ip 0}^{T(n)}
- \frac{1}{B} \sum_m \LP \OL{x_0^{(m)}} - \OL{x_\kp^{(m)}} W_{\kp \ip}^{-1(m)} V_{\ip 0}^{T(m)} \RP \RP \EL
&& \LP \OL{x_0^{(n)}} - V_{0\jp}^{(n)} W_{\jp\kp}^{-1(n)} \OL{x_\kp^{(n)}}
- \frac{1}{B} \sum_p \LP \OL{x_0^{(p)}} -  V_{0\jp}^{(p)} W_{\jp\lp}^{-1(p)} \OL{x_\lp^{(p)}}  \RP \RP \Bigg\rangle
  \\
\EEA
which, after separating diagonal and off-diagonal terms in the sums, becomes
\BEA\label{EQ_bootvar3} \LP 1-\frac{1}{B} \RP \Bigg \langle  &&
     \OL{x_0^{(n)}} \OL{x_0^{(n)}} \EL
 &-& \OL{x_0^{(n)}} \OL{x_0^{(m)}} \EL
 &-& 2 \OL{x_0^{(n)}}
    V_{0\ip}^{(n)} W_{\ip\jp}^{-1(n)} \OL{x_\kp^{(n)}} \EL
 &+& 2 \OL{x_0^{(n)}}
   V_{0\ip}^{(m)} W_{\ip\jp}^{-1(m)} \OL{x_\kp^{(m)}} \EL
 &+& \OL{x_\jp^{(n)}} W_{\jp \ip}^{-1(n)} V_{\ip 0}^{T(n)}
    V_{0\jp}^{(n)} W_{\jp\kp}^{-1(n)} \OL{x_\kp^{(n)}} \EL
 &-& \OL{x_\jp^{(n)}} W_{\jp \ip}^{-1(n)} V_{\ip 0}^{T(n)}
   V_{0\jp}^{(m)} W_{\jp\kp}^{(-1m)} \OL{x_\kp^{(m)}}
\Bigg\rangle_{n \ne m} \\
\EEA
The overall $\LP 1-\frac{1}{B} \RP$ is the expected factor for
difference between the average over the original sample and average over bootstraps.
Label the parts as in Eq.~\ref{jackvar_parts_eq0}, where now $(n)$ means the $n$'th
bootstrap resample.

For part $(a)$,
\BNE\label{EQ_bootvar_a}
\langle \OL{x_0^{(n)}} \OL{x_0^{(n)}} \rangle = \frac{1}{N^2} \sum_{ab} \LL x_0^{a(n)} x_0^{b(n)} \RR \ENE
where, in this section, the superscript $a(n)$ means the number of the data vector in the original
set that was chosen to be the $a$'th member of bootstrap resample $(n)$.
For example, if for $N=3$ our bootstrap ensemble members were members $0$, $1$ and $0$ of
the original ensemble, then $0(n)=0$, $1(n)=1$ and $2(n)=0$.
We will get contributions with nonvanishing expectation value when $a(n)=b(n)$.   If a member
of the original ensemble is chosen $m$ times in the bootstrap sample, then there will be
$m^2$ contributions.   Thus the total is the sum over all members of the original ensemble
of the square of the number of times that member was chosen for this bootstrap sample.
The probability distribution for the number of times a member appears in the bootstrap
sample is a binomial distribution  with probability $p=1/N$.
The average square of the number of times a member appears in a bootstrap resample
is just the second moment of this distribution, etc.
\BEA
 \LL (n_i) \RR &=& 1\EL
  \LL (n_i)^2 \RR &=& = 2-\OON \EL
  \LL (n_i)^3 \RR &=& =  5-\frac{6}{N}+\frac{2}{N^2} \ \ \ (N>2)\\
\EEA
Thus  the expectation value of $(a)$ is $\frac{1}{N^2} N \LP 2-\OON \RP = \frac{2}{N} -\frac{1}{N^2}$.

Part $(b)$ is the expectation value of the number of times a member was chosen in bootstrap
resample $(n)$ times the number of times it was chosen in resample $(m)$.  These two are
independent, so we get just the product of the averages, or $\frac{-1}{N}$.

For part $(c)$, break the double bar into its two components.
\BEA
(c) &=& -2\LL \OL{x_0}^{(n)} \OL{x_0 x_\ip}^{(n)} \OL{x_\ip}^{(n)} \RR
        +2\LL \OL{x_0}^{(n)} \OL{x_0}^{(n)} \OL{x_\ip}^{(n)} \OL{x_\ip}^{(n)} \RR \EL
 &=& \frac{-2}{N^3}\sum_{abc} \LL x_0^{a(n)} x_0^{b(n)}  x_\ip^{b(n)} x_\ip^{c(n)} \RR
        + \frac{2}{N^4} \sum_{abcd} \LL x_0^{a(n)} x_0^{b(n)} x_\ip^{c(n)} x_\ip^{d(n)} \RR \\
\EEA
In the first term we get a contribution when $a(n)=b(n)=c(n)$.   For each of the $N$ members
of the original ensemble we therefore get $n_a^3$ terms, where $n_a$ is the number of times
that member appeared in the bootstrap resample, so we get $N\LP 5 - \ldots \RP \LP D-P \RP$,
where the $D-P$ is from the implicit sum over $\ip$.
In the second term we get contributions when $a(n)=b(n)$ and $c(n)=d(n)$.  The probabilities
of these two conditions are not quite independent, since if one member of the original ensemble
is chosen multiple times in the bootstrap resample the other members will be chosen fewer
times.  This effect will be suppressed by a power of $\OON$, so to leading order
we just have $\LL n_a^2 \RR^2 = 4N^2 \LP D-P \RP$.  Putting in the two and overall factors of $N$ from
the left, $(c) =  \frac{-2(D-P)}{N^2} + \ldots$.

Parts $(d)$, $(e)$ and $f$ are done similarly, where to this order in $\OON$ we only need the
loop contraction in parts $(e)$ and $(f)$.

Putting it together
\BNE var_B(x_0^f) = \LP 1-\frac{1}{B} \RP \OON \LP 1 + \frac{D-P-1}{N} \RP \ENE

\section{Correcting small biases}

Once the biases in the various estimates of the error on the parameter have been
calculated, it is a simple matter to correct for them.   In particular,
we should multiply variance estimates from the derivative method by
$F_{deriv}$ in Eq.~\ref{EQ_parvar_oon_factors2}.  Note this assumes the covariance matrix was normalized
as in Eq.~\ref{EQ_covarmat_def}. For the jackknife or bootstrap done with the full sample
covariance matrix, multiply the variance by $F_{reuse}$.  This differs from
$F_{deriv}$ only in the $1$ in the denominator, the well known correction for
the difference between the sample average and the true average, which was not
included in our normalization of $C$.  For the jackknife or bootstrap analysis where a
new covariance matrix is made for each jackknife or bootstrap sample, multiply the variance
by $F_{jackknife,remake}$ or $F_{bootstrap,remake}$.
Of course, if you are rescaling error
bars instead of the variance, you should use the square root of the factor below.
(In $F_{bootstrap,remake}$ we assumed that the $\frac{B-1}{B}$
in Eq.~\ref{EQ_bootvar3} has already been accounted for.)
\BEA \label{EQ_parvar_oon_factors2}
&&F_{deriv} =  \frac{ 1 + \frac{1}{N}\LP D-P \RP + \frac{1}{N^2}\LP D-P \RP\LP D-P+2 \RP \ldots }
 { 1 - \OON\LP 1+D-P\RP + \frac{0}{N^2} } \EL
&&F_{reuse} =  \frac{ 1 + \frac{1}{N}\LP D-P \RP + \frac{1}{N^2}\LP D-P \RP\LP D-P+2 \RP \ldots }
 { 1 - \OON\LP D-P\RP + \frac{0}{N^2} } \EL
&&F_{jackknife,remake} =   1 - \frac{1}{N} \LP D-P \RP \ldots  \EL
&&F_{bootstrap,remake} =   1 + \frac{1}{N} \ldots  \\
\EEA

\section{Comparison to numerical results}

\begin{figure}[tbh]
\epsfxsize=5.5in
\epsfbox[0 0 4096 4096]{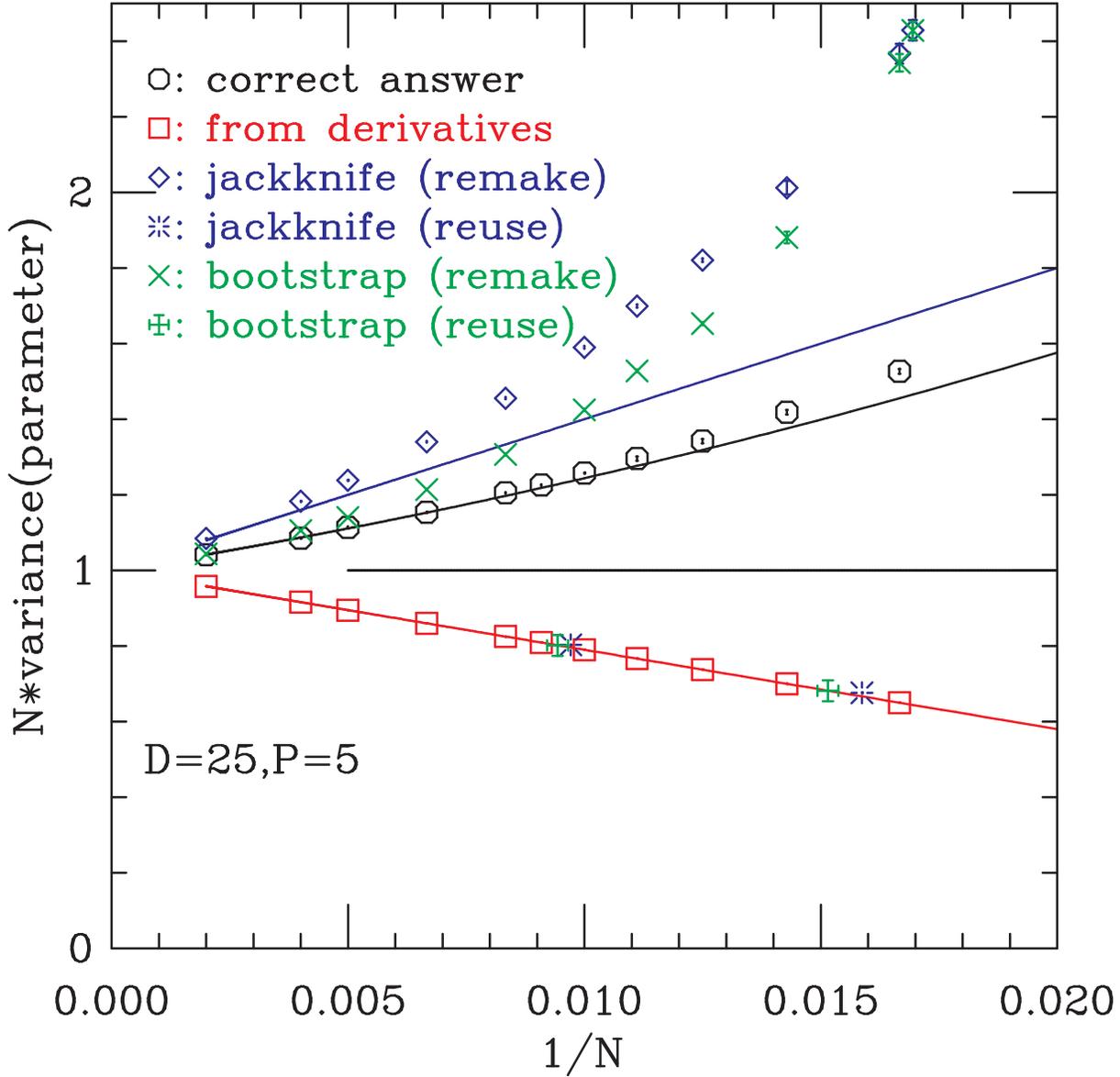}
\rule{0.0in}{0.0in}\vspace{1.0in}\\
\caption{
\label{FIG_sim2}
Numerical results from Fig.\protect\ref{FIG_sim1} together with
the order $\OON$ and $\frac{1}{N^2}$ results from the previous section.
The meaning of the symbols is the same as in Fig.~\protect\ref{FIG_sim1}.
}
\end{figure}

\begin{figure}[tbh]
\epsfxsize=5.5in
\epsfbox[0 0 4096 4096]{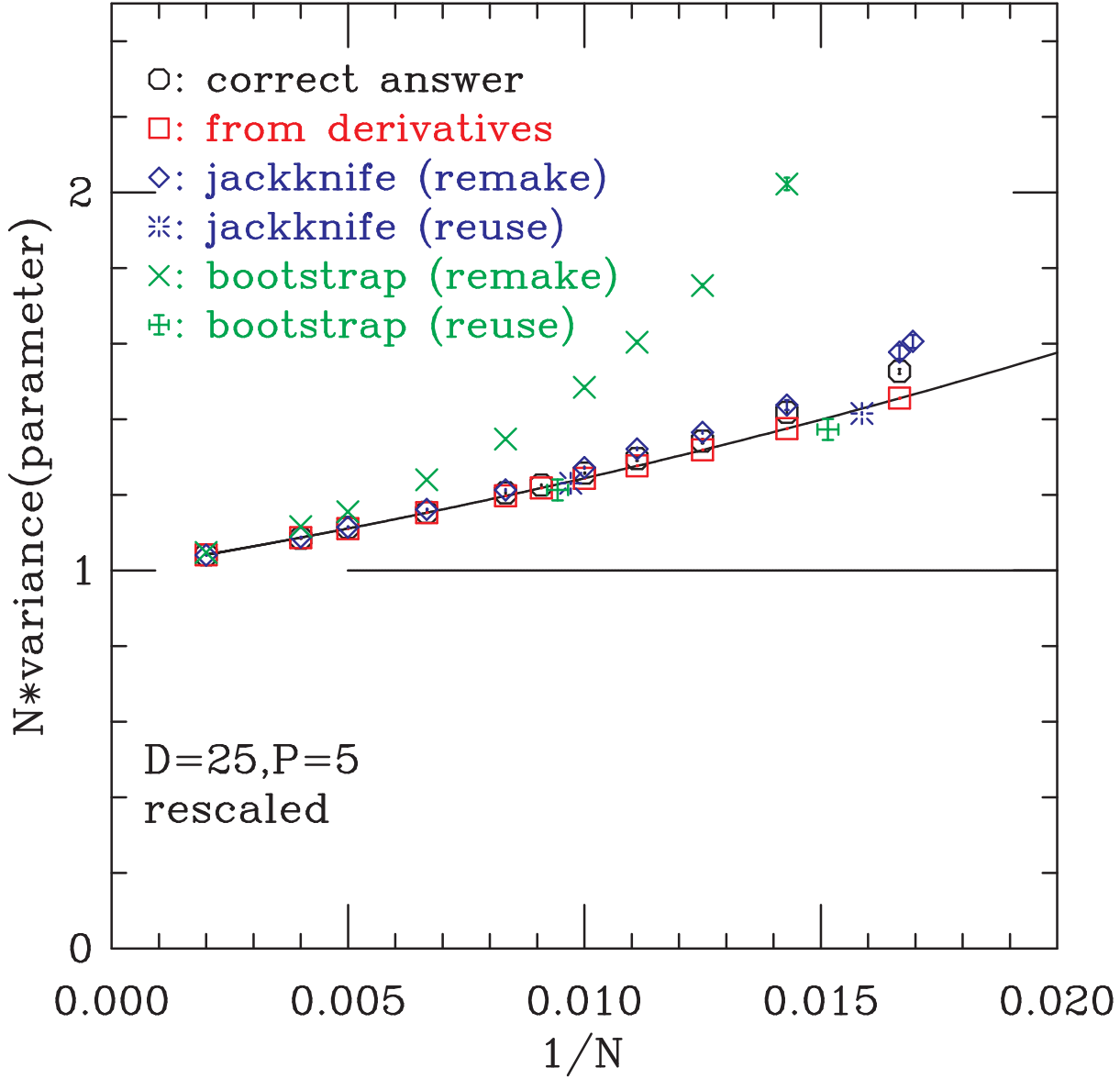}
\rule{0.0in}{0.0in}\vspace{1.0in}\\
\caption{
\label{FIG_sim3}
Numerical results from Fig.\protect\ref{FIG_sim1} corrected for bias
up to corrections of order $\frac{1}{N^2}$ for the jackknife with remade
covariance matrices and order $\frac{1}{N^3}$ for the methods with fixed
covariance matrix.
}
\end{figure}

In Fig.~\ref{FIG_sim2} we plot the order $\OON$ forms for the variance
of the parameter and the various methods of estimating it together with
the numerical data.  The horizontal axis has been inverted to $\OON$.
Figure~\ref{FIG_sim3} shows the same data, with the estimates for the
variance corrected for bias (up to errors of order $\frac{1}{N^3}$ or $\frac{1}{N^2}$).
Here the lines for the actual variance of the parameter (black)
and for the derivative or resampling with the full sample
covariance matrix (red) are second order in $\OON$, while
the line for the jackknife with remade covariance matrices (blue)
is only first order in $\OON$.
As an aside, we note that although the lowest order corrections for the bootstrap
with remade covariance matrices are smaller than for the other methods,
the next order corrections appear to be larger.


\section*{Appendix I}

Since estimating the goodness of fit is as important as estimating the errors
on the parameters, we quote some results here.  Note that what we call $\chi^2$
(with the covariance matrix estimated from our data) is more properly called
$T^2$, but we stick with the common usage in the lattice gauge community.

The probability
distribution for $\chi^2$ is known\cite{Morrison}.  In terms of $N$ and $d$,

\BNE\label{Fdist1_eq} Prob(\chi^2) = \frac{N^{-d/2} \Gamma(N/2)}{\Gamma(d/2)\Gamma((N-d)/2)}\,
(\chi^2)^{(d-2)/2} \LP 1 + \frac{1}{N}\chi^2 \RP^{-N/2} \ENE
We can compare to the $\chi^2$ distribution:
\BE Prob(\chi^2) = C\, (\chi^2)^{(d-2)/2} e^{-\chi^2/2} \EE
and see that in the limit of large $N$ they are the same.

From moments of Eq.~\ref{Fdist1_eq} we see that the mean and variance of
$\chi^2$ depend on the sample size.
Using
\BNE I(D,N) \equiv \int_0^\infty d(\chi^2) (\chi^2)^{(D-2)/2} \LP 1 + \frac{1}{N}\chi^2 \RP^{-N/2}
= \frac{N^{D/2} \Gamma(D/2)\Gamma((N-D)/2)}{\Gamma(N/2)} \ENE,
\BNE \LL \chi^2 \RR = \frac{I(D+2,N)}{I(D,N)} = \frac{N\frac{D}{2}}{\frac{N-D-2}{2}} =
\frac{D}{1-\frac{D+2}{N}} \ENE
\BNE \LL (\chi^2)^2 \RR = \frac{I(D+4,N)}{I(D,N)} = \frac{N^2\frac{D}{2}\LP\frac{D+2}{2}\RP }
{ \LP\frac{N-D-2}{2} \RP \LP\frac{N-D-4}{2} \RP } =
\frac{D(D+2) }{ \LP 1-\frac{D+2}{N} \RP \LP 1-\frac{D+4}{N} \RP } \ENE
(Note this is using our normalization of the covariance matrix).

Taking the connected part, or variance, and expanding in $\OON$, this is
\BNE \label{EQ_varchisq_1} var(\chi^2) = 2\dof \LP 1 + \frac{3\dof+6}{N} \RP \ENE

Estimates of confidence levels, or probability (over trials) that $\chi^2$
would exceed the value in your experiment, can be found by integrating
Eq.~\ref{Fdist1_eq}.

%

\section*{Appendix II}

The customary estimate for the variance of the parameters, or from jackknife or bootstrap
resamplings with the covariance matrix held fixed, has zero coefficient at the next order.
\BEA N var(x_0^f)_{derivative}  &=&  A_{00}^{-1} \EL
 &=& U_{00} - V_{0\kp} W_{\kp \lp}^{-1} V_{\lp 0}^T \EL
&=& \OON \LP \delta_{00} + \OLL{x_0 x_0} \RP \EL
&-& \OON \LP \OLL{x_0 x_\kp} \RP \LP \delta_{\kp \lp } - \OLL{x_\kp x_\lp}
  + \OLL{x_\kp x_\mp} \OLL{x_\mp x_\lp} \ldots \RP \OLL{x_\lp x_{0}} \\
\EEA
Again, we only need two loop contractions from the terms with eight $x$'s.
\BEA
 N\, varest(p_0) &=& \delta_{00} \label{varest_p0_eq} \\
 &+& \OLL{ x_0 x_0}(12) \EL
 &-& \OLL{ x_0 x_\kp} \OLL{ x_\kp x_0} (14)(23) \EL
 &+& \OLL{ x_0 x_\kp} \OLL{ x_\kp x_\lp} \OLL{ x_\lp x_0} (16)(23)(45) \EL
 &+& \OLL{ x_0 x_\kp} \OLL{ x_\kp x_\lp} \OLL{ x_\lp x_0} (16)(24)(35) \EL
 &+& \OLL{ x_0 x_\kp} \OLL{ x_\kp x_\lp} \OLL{ x_\lp x_0} (16)(25)(34) \EL
 &-& \OLL{ x_0 x_\kp} \OLL{ x_\kp x_\lp} \OLL{ x_\lp x_\mp} \OLL{ x_\mp x_0} (18)(27)(35)(46) \EL
 &-& \OLL{ x_0 x_\kp} \OLL{ x_\kp x_\lp} \OLL{ x_\lp x_\mp} \OLL{ x_\mp x_0} (18)(27)(36)(45) \EL
 &=& 1 - \OON - \frac{N-1}{N^2}(D-P) \EL
 &+& \frac{N-1}{N^3} (D-P)^2 
 + \frac{N-1}{N^3} (D-P) \EL
 &+& \frac{N-1}{N^3} (-1)(D-P) 
 - \frac{(N-1)^2}{N^4} (D-P) \EL
 &-& \frac{(N-1)^2}{N^4} (D-P)^2 \EL
 &=& 1 - \OON\LP 1+D-P\RP + \frac{0}{N^2} \\
\EEA
If $D=P=1$ this is just $ \OON \LL 1+\OLL{x_1 x_1}\RR = 1 - \OON $ as it must be.

\end{document}